\documentclass[aps,prl,superscriptaddress,twocolumn,longbibiliography]{revtex4-2}

\usepackage[hidelinks]{hyperref}
\hypersetup{
	colorlinks,
	linkcolor={red},
	citecolor={blue},
	urlcolor={blue}
}
\usepackage{physics}
\usepackage{balance}
\usepackage{amssymb}
\usepackage{suffix}
\usepackage{mathtools}
\usepackage[utf8]{inputenc}
\usepackage{booktabs}
\usepackage{cases}
\usepackage[multiple]{footmisc}
\usepackage{dcolumn}
\usepackage{color,soul}
\usepackage{rotating}
\usepackage{perpage}
\usepackage{siunitx}
\usepackage{xcolor}
\usepackage{soul}
\usepackage{amsmath}
\usepackage{tikz}
\usepackage[T1]{fontenc}
\usepackage{etoolbox}
\usepackage{graphics}
\usepackage{siunitx}
\usepackage{float}	
\usepackage{collref}
\usepackage{multirow}
\usepackage{mathtools}
\usepackage{bm}
\usepackage{url}

\usepackage{tikz}
\usepackage{tikz-3dplot}
\usepackage{accents}
\usepackage{verbatim}

\makeatletter
\newcommand{\doublewidetilde}[1]{{%
		\mathpalette\double@widetilde{#1}%
}}
\newcommand{\double@widetilde}[2]{%
	\sbox\z@{$\m@th#1\widetilde{#2}$}%
	\ht\z@=.9\ht\z@
	\widetilde{\box\z@}%
}
\makeatother

\usepackage{etoolbox,lipsum}

\begin{document}
	
	\title{Interplay of nonrelativistic and relativistic spin splittings in altermagnets}

	\author{Daegeun~Jo}
	\email{daegeun.jo@physics.uu.se}
	\affiliation{Department of Physics and Astronomy, Uppsala University, P.O. Box 516, SE-75120 Uppsala, Sweden}
	\affiliation{Wallenberg Initiative Materials Science for Sustainability, Uppsala University, SE-75120 Uppsala, Sweden}
	
	\author{Peter~M.~Oppeneer}
	\email{peter.oppeneer@physics.uu.se}
	\affiliation{Department of Physics and Astronomy, Uppsala University, P.O. Box 516, SE-75120 Uppsala, Sweden}
	\affiliation{Wallenberg Initiative Materials Science for Sustainability, Uppsala University, SE-75120 Uppsala, Sweden}

	\author{Mohsen~Yarmohammadi}
	\email{mohsen.yarmohammadi@georgetown.edu}
	\affiliation{Department of Physics, Georgetown University, Washington DC 20057, USA}
	
	\date{\today}

	\begin{abstract}
		Despite their compensated magnetic moments, altermagnets~(AMs) exhibit nonrelativistic spin splitting (NRSS) that offers routes to spintronic functionality without relying on relativistic spin-orbit coupling. However, how NRSS influences relativistic phenomena remains largely unexplored. Here we show that NRSS plays a crucial role in reshaping the Rashba effect in AMs with broken inversion symmetry. Using a tight-binding model, we demonstrate that Rashba spin splitting, whose magnitude is typically limited by the strength of spin-orbit coupling, is governed by NRSS energy scales. The resulting Rashba bands combined with NRSS generate anomalously large charge-to-spin conversion with a Néel-vector-tunable spin polarization. First-principles calculations for the noncentrosymmetric AM GdAlSi corroborate the emergence of these effects in a real system. Our results reveal the interplay between nonrelativistic and relativistic effects in AMs and identify them as fertile platforms for spin-orbitronic applications.
	\end{abstract}
	
	\maketitle
	{\allowdisplaybreaks
		\textit{Introduction}---Altermagnets (AMs) are unconventional magnetic materials that exhibit compensated spin ordering in real space but display nonrelativistic spin splitting (NRSS) in momentum space~\cite{hayami2019momentum, yuan2020giant,smejkal2022beyond, smejkal2022emerging, mazin2022, bai2024altermagnetism}. Although AMs have attracted significant interest owing to their NRSS~\cite{gonzalez2021efficient, bai2022observation, karube2022observation, lee2024broken, fedchenko2024observation, krempasky2024altermagnetic, osumi2024observation}, a natural question is whether spin-orbit coupling (SOC) can further give rise to intriguing relativistic phenomena relevant to spin-orbitronic applications~\cite{manchon2015new, manchon2019current, trier2022oxide}. Indeed, recent studies have explored various SOC-induced effects in AMs, including the anomalous Hall effect~\cite{smejkal2020crystal, feng2022anomalous, gonzalez2023spontaneous, takagi2025spontaneous}, magneto-optic effects~\cite{zhou2021crystal, wang2022magneto, pan2026experimental, luo2026symmetry}, and x-ray magnetic circular dichroism~\cite{lovesey2023templates, hariki2024xray, amin2024nanoscale, galindez2025revealing}. 
		
		The spin Rashba effect (SRE) is another key relativistic phenomenon, in which broken inversion symmetry and SOC lift spin degeneracy and generate momentum-dependent spin textures~\cite{bychkov1984, manchon2015new, bihlmayer2022rashba}. Maximizing Rashba spin splitting is a central goal in materials science and spin-orbitronics, but its magnitude is usually limited by the strength of SOC~
        \cite{sunko2017asymmetry}. The SRE provides versatile functionalities in spin-orbitronics when combined with ferromagnetic~\cite{chernyshov2009evidence, miron2010current, fang2011spin, miron2011perpendicular, kurebayashi2014antidamping, ciccarelli2016room} or antiferromagnetic~\cite{zelezny2014relativistic, wadley2016electrical, bodnar2018writing, olejnik2018terahertz, salemi2019orbitally} order. Studies of the SRE have recently been extended to AMs~\cite{rao2024tunable, amundsen2024rkky, ezawa2024intrinsic, chen2025helicity, mukasa2025finite, yarmohammadi2025anisotropic, kapri2025spin, trama2025nonlinear, chien2025electrically, duan2026neel, yarmohammadi2026spin, yarmohammadi2026slow, yarmohammadi2026efficient, yarmohammadi2026floquet, yarmohammadi2026giant}, yet a microscopic and general understanding of how Rashba physics is modified in such systems remains largely unexplored. 
		
		In this Letter, we show that AMs with broken inversion symmetry exhibit distinctive Rashba physics through the interplay of nonrelativistic and relativistic effects. Due to NRSS, weak SOC dresses the already spin-split bands with Rashba spin textures. The Rashba spin splitting is therefore governed by nonrelativistic energy scales, which can be much larger than SOC. Accordingly, charge-to-spin conversion through the spin Edelstein effect~\cite{edelstein1990spin} can be significantly enhanced or modified. We establish this mechanism using a tight-binding model and further corroborate it through first-principles calculations for the noncentrosymmetric AM GdAlSi. Our findings highlight the key role of NRSS in relativistic Rashba physics and suggest that Rashba AMs enable efficient charge-to-spin conversion, making them promising platforms for magnetic device applications. 
		\begin{figure}[t]
			\center\includegraphics[width=0.95\columnwidth]{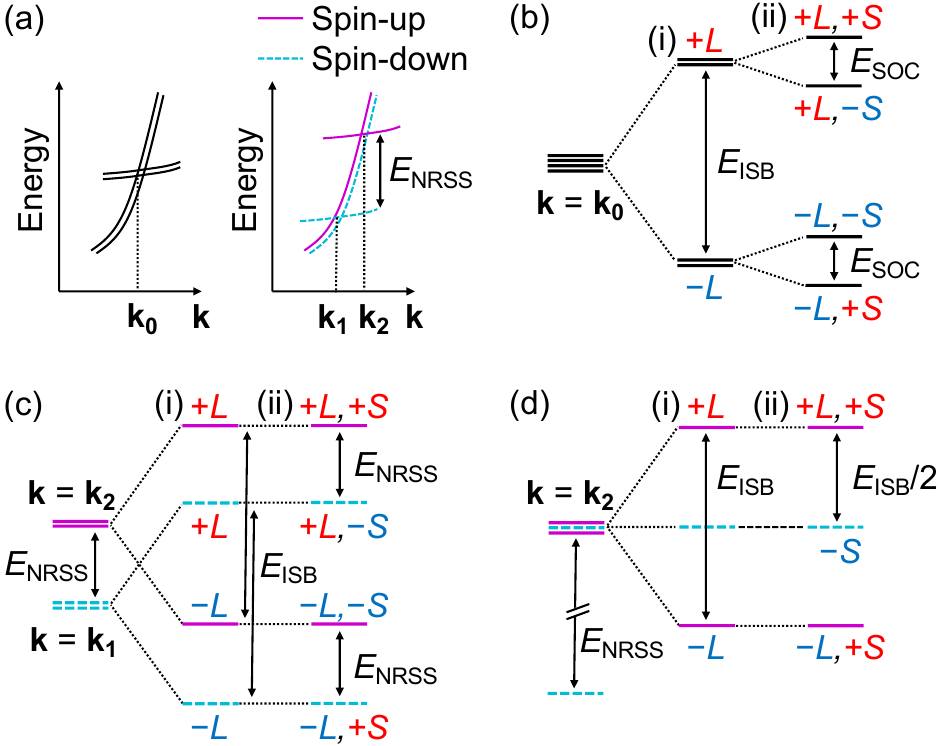}
			\caption{(a) Schematic illustration of crossing bands in inversion-symmetric systems without (left) and with (right) NRSS. (b) Rashba splitting of the spin-degenerate states due to ISB and SOC, which are denoted by (i) and (ii), respectively, when $E_\mathrm{ISB} \gg E_\mathrm{SOC}$. For each state, $\pm L$ and $\pm S$ represent the helicity of orbital angular momentum and spin textures, respectively. (c),(d) Rashba splitting with NRSS, when (c) $E_\mathrm{NRSS} \lesssim E_\mathrm{ISB}$ and (d) $ E_\mathrm{NRSS} \gg E_\mathrm{ISB}$.
			}
			\label{fig_schematic} 
		\end{figure}

		\textit{Schematic picture}---We first summarize the central idea of our work. The left panel of Fig.~\ref{fig_schematic}(a) schematically shows the band structure of a centrosymmetric nonmagnet or antiferromagnet, where combined space-inversion and time-reversal ($\mathcal{PT}$) symmetry enforces double degeneracy at every crystal momentum $\mathbf{k}$. Two different orbital bands cross at $\mathbf{k} = \mathbf{k}_0$, forming a four-fold degeneracy. Figure~\ref{fig_schematic}(b) illustrates how inversion symmetry breaking (ISB) and weak SOC lift this degeneracy through Rashba effects~\cite{sunko2017asymmetry,bihlmayer2022rashba}. ISB first induces the orbital Rashba effect (ORE)~\cite{park2011orbital, park2013orbital, go2017toward}, generating $\mathbf{k}$-dependent orbital angular momentum (OAM) textures with opposite helicities, $\pm L$, and opening an energy gap $E_{\mathrm{ISB}}$. Since this ORE arises from spin-independent orbital hybridization, the orbital Rashba bands remain spin-degenerate in the absence of SOC. Subsequently, weak SOC leads to the SRE by transferring the Rashba OAM texture to the spin texture, with helicities  $\pm S$, and lifting spin degeneracy on the purely relativistic energy scale $E_{\mathrm{SOC}}$. 
		
		Next, we consider AMs in which broken $\mathcal{PT}$ symmetry already allows NRSS to lift the spin degeneracy. For simplicity, as shown in the right panel of Fig.~\ref{fig_schematic}(a), we assume that one band remains spin-degenerate and crosses the two spin-split bands at $\mathbf{k} = \mathbf{k}_1$ and $\mathbf{k} = \mathbf{k}_2$. Figure~\ref{fig_schematic}(c) illustrates the regime of moderate NRSS, $E_\mathrm{SOC} < E_\mathrm{NRSS} \lesssim E_\mathrm{ISB}$, in which $\mathbf{k}_1$ and $\mathbf{k}_2$ are sufficiently close. Because ISB alone does not mix opposite spins, the four interacting states at these crossings can be viewed approximately as the two spin-up states near $\mathbf{k}_2$ and the two spin-down states near $\mathbf{k}_1$. ISB induces an ORE independently in the two spin channels, since each spin-polarized band selectively hybridizes within the corresponding spin channel. As a result, the orbital Rashba bands become spin-polarized, inheriting the NRSS of the AM through ISB. Weak SOC then produces a Rashba spin texture similar to that in Fig.~\ref{fig_schematic}(b), while the resulting spin splitting is governed by $E_\mathrm{NRSS}$ rather than by $E_\mathrm{SOC}$. Hence, the Rashba spin splitting can be substantially enhanced by NRSS.
		
		A distinct regime emerges when $E_\mathrm{NRSS} \gg E_\mathrm{ISB}$ [Fig.~\ref{fig_schematic}(d)]. Near $\mathbf{k} = \mathbf{k}_2$, the lower spin-down state is far from the crossing; thus, ISB induces an ORE predominantly within the spin-up channel, while the spin-down bands remain largely unaffected. In this regime, the spin splitting is maximized, reaching $\sim E_\mathrm{ISB}/2$. However, the SOC-induced hybridization is suppressed as the opposite-spin bands are more energetically separated. With increasing $E_\mathrm{NRSS}$, the Rashba spin texture becomes weaker, while the spin polarization along the N\'eel vector becomes more pronounced. AMs thus enrich Rashba physics by introducing the additional scale $E_\mathrm{NRSS}$ into the conventional interplay between $E_\mathrm{ISB}$ and $E_\mathrm{SOC}$~\cite{sunko2017asymmetry}. 
		
		\textit{Tight-binding model}---To illustrate the above discussion, we consider a multi-orbital tight-binding model for a two-dimensional $d$-wave AM with inversion symmetry~\cite{jo2025weak}. As shown in Fig.~\ref{fig_tb}(a), the square lattice contains two sublattices $A$ and $B$ with opposite spins, each hosting three orbitals, $d_{xy}$, $d_{yz}$, and $d_{zx}$. The Hamiltonian is expressed in the tensor-product space of orbital, sublattice, and spin degrees of freedom~\cite{supp}:
		\begin{equation}\label{eq:H_alter}
			\hat{H}_\mathrm{AM}(\mathbf{k}) = \hat{h}(\mathbf{k}) \hat{\tau}_x + \frac{\Delta}{2} (\hat{\ell}_y^2 - \hat{\ell}_x^2) \hat{\tau}_z  - J \hat{\tau}_z  \hat{\mathbf{n}} \cdot \hat{\bm{\sigma}} + \lambda \hat{\bm{\ell}} \cdot \hat{\bm{\sigma}},
		\end{equation}
		where $\hat{\ell}_i$ ($i=x,y,z$) are the dimensionless OAM operators (in units of $\hbar$) in the orbital basis, and $\tau_i$ and $\sigma_i$ are Pauli matrices acting in the sublattice and spin subspaces, respectively. Identity operators are implicit in subspaces where no operator is shown explicitly.

		\begin{figure}[t]
			\center\includegraphics[width=0.95\columnwidth]{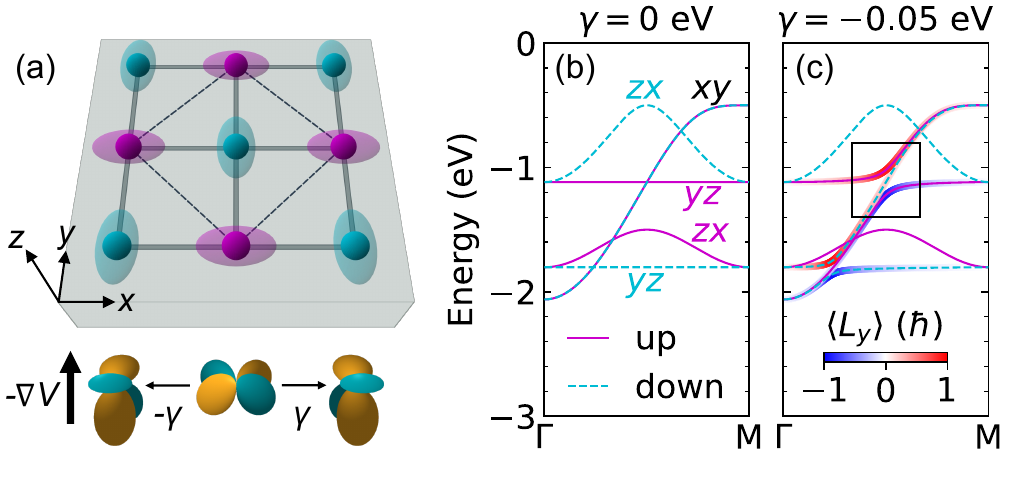}
			\caption{(a) Schematics of the tight-binding model. The dashed square denotes the unit cell with lattice constant $a=5$~\AA, consisting of spin-up (magenta) and spin-down (cyan) sublattices. Ellipses represent the anisotropic crystal field. Introducing ISB generates an electric field $-\nabla V $, which deforms the orbitals and induces an antisymmetric hopping $\gamma$. (b),(c) Band structures without SOC for (b) $\gamma = 0$~eV and (c) $\gamma = -0.05$~eV along $\Gamma=(0,0)$ to $\mathrm{M} = (\frac{\sqrt{2}\pi}{a},0)$. The $d$-orbital character for each band is shown in (b), and the expectation value of the OAM is indicated by color in (c).
			}
			\label{fig_tb} 
		\end{figure}
		
		The first term on the right-hand side of Eq.~\eqref{eq:H_alter} describes nearest-neighbor hopping between sublattices. For simplicity, we retain only $\pi$-bonding hopping, thus the hopping matrix in the orbital basis is given by $\hat{h}(\mathbf{k}) = -2t \, \mathrm{diag}[\cos(k_xa) + \cos(k_ya),  \cos(k_ya), \cos(k_xa)]$, where $t = 0.5$~eV. The second term represents the sublattice-dependent crystal field splitting $\pm \Delta$ between the $d_{yz}$ and $d_{zx}$ orbitals, reflecting the $C_4$-related anisotropy. It also describes an antiferroic electric quadrupole order, since $\hat{\ell}_y^2 - \hat{\ell}_x^2$ corresponds to a quadrupole operator. The third term describes antiferromagnetic spin ordering with the Néel vector along $\hat{\mathbf{n}}$ and exchange parameter $J$. Unless otherwise specified, we fix $\hat{\mathbf{n}} = \hat{\mathbf{z}}$ and $J=0.5$~eV throughout the paper. The second and third terms together make the system a $d$-wave AM, naturally giving rise to a magnetic octupole order~\cite{bhowal2024ferroically}. The last term describes SOC with strength $\lambda$. ISB is not included at this stage.

		Figure~\ref{fig_tb}(b) shows the band structure in the absence of SOC ($\lambda=0$~eV). The magnitude of the NRSS is determined by the crystal field splitting $\Delta$ between the $d_{yz}$ and $d_{zx}$ orbitals, which is set to 2~eV here. We display only the lower six bands, as the upper six bands are nearly symmetric about 0~eV. The band with $d_{xy}$ orbital character remains spin-degenerate; it preserves $C_4$ symmetry and its energy is therefore sublattice-independent.

		Next, we introduce ISB, which leads to the ORE, while still neglecting SOC. We assume that ISB---arising from structural asymmetry or an applied gate voltage---produces an electric field $-\nabla V$ along $\hat{\mathbf{z}}$. This induces a deformation of the orbital wave functions, resulting in antisymmetric interorbital hopping~[Fig.~\ref{fig_tb}(a)], $\langle d_{xy} \vert \hat{H}_\mathrm{ISB} \vert d_{yz} ( \pm \frac{a}{\sqrt{2}} \hat{\mathbf{x}}  ) \rangle = \langle d_{xy} \vert \hat{H}_\mathrm{ISB} \vert d_{zx} ( \pm \frac{a}{\sqrt{2}} \hat{\mathbf{y}} ) \rangle \equiv \pm \gamma$ \cite{park2013orbital, zhong2013theory, khalsa2013theory}. In momentum space, this hopping yields
		\begin{equation}\label{eq:H_ISB}
			\hat{H}_\mathrm{ISB}(\mathbf{k}) = -2 \gamma \left[\sin (\frac{k_x a}{\sqrt{2}}) \hat{\ell}_y - \sin (\frac{k_y a}{\sqrt{2}}) \hat{\ell}_x \right] \hat{\tau}_x,
		\end{equation}
		which implies OAM-momentum locking and reduces near the $\Gamma$ point to the standard ORE form, $\alpha_\mathrm{OR} \hat{\bm{\ell}} \cdot (\hat{\mathbf{z}} \times \mathbf{k})$~\cite{park2011orbital, park2013orbital, go2017toward}, with the orbital Rashba constant $\alpha_\mathrm{OR} = \sqrt{2} a \gamma$. 
		
		Figure~\ref{fig_tb}(c) shows the band structure for the Hamiltonian $\hat{H}(\mathbf{k}) = \hat{H}_\mathrm{AM}(\mathbf{k}) + \hat{H}_\mathrm{ISB}(\mathbf{k})$ with $\gamma=-0.05$~eV. The spin-up and spin-down $d_{yz}$ bands selectively couple to the corresponding spin channels of the $d_{xy}$ band, opening an orbital Rashba gap in each spin channel. To reveal the OAM-momentum locking, we compute the expectation value of the OAM operator, $\hat{\mathbf{L}} = \hbar \hat{\bm{\ell}}$, for each state. For $\mathbf{k} = (k_x, 0)$, the OAM is polarized along $\hat{\mathbf{y}}$, as shown in Fig.~\ref{fig_tb}(c), demonstrating that AMs with ISB host spin-polarized orbital Rashba bands. 
		
		\textit{Rashba spin splitting enhanced by NRSS}---We now include SOC to investigate the SRE. In the presence of SOC, the magnetic point group depends on the direction of $\hat{\mathbf{n}}$. Starting from the crystallographic point group $4/mmm$ of $\hat{H}_\mathrm{AM}(\mathbf{k})$, out-of-plane ($\hat{\mathbf{n}} = \hat{\mathbf{z}}$) and in-plane ($\hat{\mathbf{n}} = \hat{\mathbf{y}}$) Néel vectors lead to the magnetic point groups $4'/mm'm$ and $m'mm'$, respectively. The former forbids net magnetization, while the latter allows weak ferromagnetism~\cite{jo2025weak, cheong2024altermagnetism}. ISB described by Eq.~\eqref{eq:H_ISB} further reduces these magnetic point groups to $4'm'm$ and $m'm2'$, respectively. Because the macroscopic magnetization remains zero for $\hat{\mathbf{n}} = \hat{\mathbf{z}}$, we focus exclusively on this configuration to demonstrate that the enhanced Rashba phenomena discussed herein do not rely on net magnetization.

		Taking $\lambda=0.02$~eV as a representative weak SOC strength, which yields a small spin Rashba coupling constant $\alpha_\mathrm{SR} \sim \vert \sqrt{2} a \gamma \lambda / t \vert \approx 14$~meV~\r{A}~\cite{zhong2013theory, khalsa2013theory}, we examine the spin textures near the orbital Rashba splitting for three values of $\Delta$. First, the conventional antiferromagnetic case without NRSS [Fig.~\ref{fig_texture}(a), $\Delta=0$~eV] exhibits the usual Rashba spin texture, where the spin is locked to $\pm \hat{\mathbf{y}}$ for $\mathbf{k}=(k_x,0)$. The Rashba spin splitting $\sim 2\lambda$ is determined by SOC. In contrast, in an AM with moderate NRSS [Fig.~\ref{fig_texture}(b), $\Delta=0.2$~eV], the Rashba spin texture is induced on bands already spin-polarized along $\hat{\mathbf{n}} = \hat{\mathbf{z}}$. The Rashba spin splitting is therefore governed by NRSS. As NRSS further increases [Fig.~\ref{fig_texture}(c), $\Delta=2$~eV], the magnitude of the spin splitting is maximized and bounded by the orbital Rashba splitting, which is set by $\gamma$ and is typically much larger than the relativistic spin splitting. The results for these three cases are consistent with the schematic picture in Figs.~\ref{fig_schematic}(b)--(d).
		
		The main features of the SRE in AMs are more clearly seen in Figs.~\ref{fig_texture}(d)--(f), which show the spin textures on the Fermi contours at an energy $-1.0$~eV for the cases of Figs.~\ref{fig_texture}(a)--(c), respectively. In Fig.~\ref{fig_texture}(d), the Fermi contours with opposite spin helicities mostly overlap owing to the weak relativistic Rashba splitting. Once NRSS is introduced [Figs.~\ref{fig_texture}(e) and \ref{fig_texture}(f)], these pairs of Fermi contours become strongly asymmetric, with their splitting governed by the larger nonrelativistic energy scales. Note that the in-plane Rashba spin texture becomes weaker for larger NRSS, since the spin polarization tends to align more strongly with $\hat{\mathbf{n}}=\hat{\mathbf{z}}$. Nevertheless, this interplay between the relativistic Rashba texture and the NRSS-induced spin polarization can significantly enhance or modify magnetoelectric responses, as shown below. 
		
		\begin{figure}[t]
			\center\includegraphics[width=1\columnwidth]{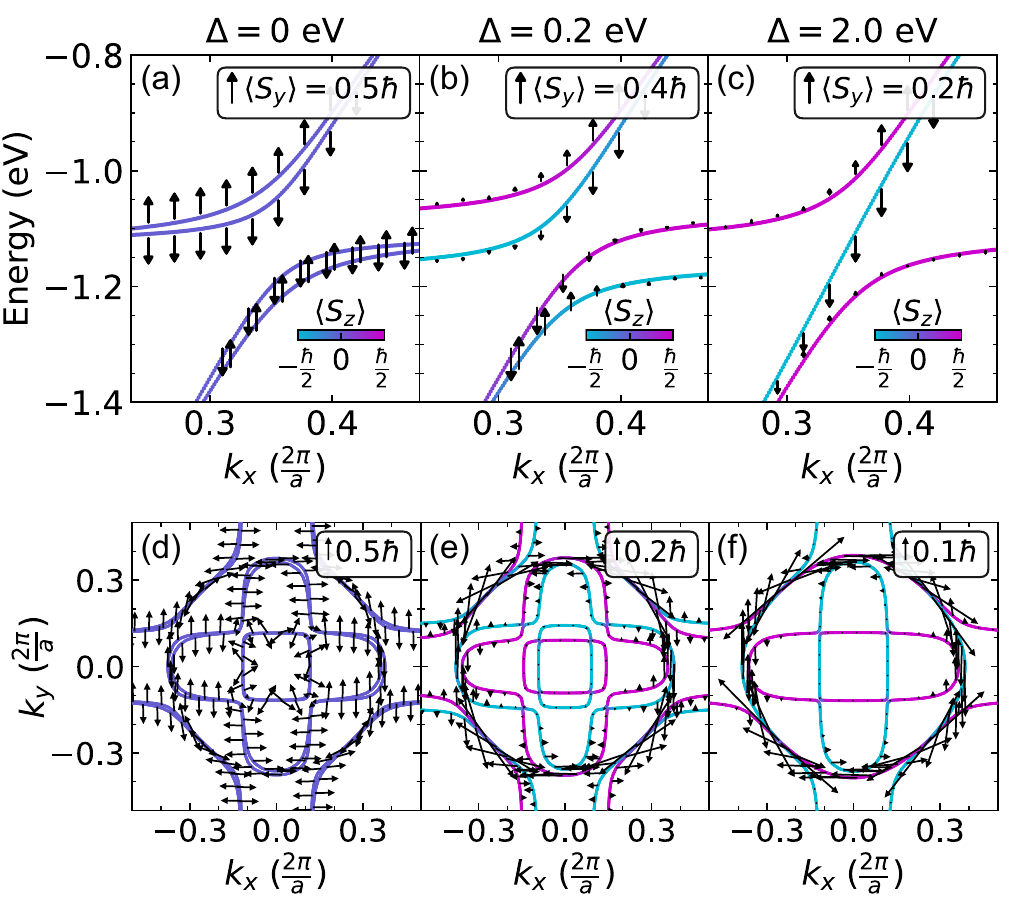}
			\caption{(a)--(c) Spin textures in the boxed region of Fig.~\ref{fig_tb}(c) with SOC ($\lambda = 0.02$~eV), for different NRSS strengths controlled by (a) $\Delta=0$~eV, (b) $\Delta=0.2$~eV, and (c) $\Delta=2.0$~eV. Arrows and colors indicate the spin expectation values along the $y$ axis and the Néel vector ($z$ axis), respectively. (d)--(f) Spin textures on the Fermi contours at energy $-1.0$~eV for the systems in (a)--(c), respectively. 
			}
			\label{fig_texture} 
		\end{figure}
		
		\textit{Spin Edelstein effect}---NRSS can strongly enhance the spin Rashba--Edelstein effect (SREE)~\cite{edelstein1990spin, bihlmayer2022rashba}. In the intraband mechanism of the SREE, an applied electric field $\mathbf{E}$ shifts the Fermi surface with a Rashba spin texture, generating a nonequilibrium spin polarization $\delta \mathbf{S} \propto \hat{\mathbf{z}} \times \mathbf{E}$, where $\hat{\mathbf{z}}$ denotes the direction of the internal electric field associated with ISB. In conventional Rashba systems with weak SOC, the two Fermi surfaces with opposite spin helicities are often nearly identical [Fig.~\ref{fig_texture}(d)], so their contributions largely cancel~\cite{johansson2021spin}. In Rashba AMs, by contrast, the Fermi surfaces can be strongly asymmetric [Figs.~\ref{fig_texture}(e) and \ref{fig_texture}(f)], reducing their mutual compensation and significantly enhancing the SREE. 
		
		The spin-polarized nature of the Rashba bands additionally gives rise to unconventional magnetoelectric responses. In Rashba ferromagnets with magnetization along $\hat{\mathbf{m}}$, the SREE-induced $\delta \mathbf{S} \propto \hat{\mathbf{z}} \times \mathbf{E}$ leads to an interband spin-torque mechanism through the exchange interaction, generating an additional nonequilibrium spin polarization $\delta \mathbf{S} \propto \hat{\mathbf{m}} \times (\hat{\mathbf{z}} \times \mathbf{E})$~\cite{miron2011perpendicular, kurebayashi2014antidamping, ciccarelli2016room, manchon2008theory, li2015intraband}. Because the Rashba bands in AMs are strongly spin-polarized along $\hat{\mathbf{n}}$, an electric field can similarly induce $\delta \mathbf{S} \propto \hat{\mathbf{n}} \times (\hat{\mathbf{z}} \times \mathbf{E})$. Consequently, Rashba AMs can exhibit ferromagnet-like responses through $\mathbf{k}$-dependent spin splitting, even though net magnetization is forbidden by symmetry. This contrasts with other spin-orbit phenomena, such as the anomalous Hall effect, whose symmetry requirement coincides with that for weak ferromagnetism~\cite{mcclarty2024landau}.

		To investigate these effects, we numerically evaluate the current-induced spin polarization $\delta \mathbf{S} = \delta \mathbf{S}^\mathrm{intra} + \delta \mathbf{S}^\mathrm{inter}$ for $\mathbf{E} = E_x \hat{\mathbf{x}}$, where $\delta \mathbf{S}^\mathrm{intra}$ and $\delta \mathbf{S}^\mathrm{inter}$ are the intraband and interband contributions, respectively, given by~\cite{Salemi2021}
		%
		{\small\begin{subequations}
				\begin{align}
					\delta \mathbf{S}^\mathrm{intra} = &
					\frac{e\hbar E_x}{2 N_\mathbf{k} \Gamma}\sum_{\mathbf{k},n} 
					(\frac{\partial f_{n\mathbf{k}}}{\partial \varepsilon_{n\mathbf{k}}} ) \langle \psi_{n\mathbf{k}} \vert \hat{\mathbf{S}}  \vert \psi_ {n\mathbf{k}}\rangle 
		\langle \psi_{n\mathbf{k}} \vert \hat{v}_x \vert \psi_{n\mathbf{k}} \rangle , \label{eq:kubo_surf} \\
					\delta \mathbf{S}^\mathrm{inter}  = &
					-\frac{e\hbar E_x}{N_\mathbf{k}}  \sum_{\mathbf{k}, n \neq m} 
					(f_{n\mathbf{k}} - f_{m\mathbf{k}}) \nonumber  \\ 
					&  \times  \Im\left[
					\frac
					{\langle \psi_{n\mathbf{k}} \vert \hat{\mathbf{S}}  \vert \psi_ {m\mathbf{k}}\rangle 
						\langle \psi_{m\mathbf{k}} \vert \hat{v}_x \vert \psi_{n\mathbf{k}} \rangle 
					}
					{(\varepsilon_{n\mathbf{k}} - \varepsilon_{m\mathbf{k}})( \varepsilon_{n\mathbf{k}} - \varepsilon_{m\mathbf{k}} + i\Gamma )} \right]. \label{eq:kubo_sea}
				\end{align}
		\end{subequations}}Here, $e$ is the elementary charge, $N_\mathbf{k}$ is the number of $\mathbf{k}$ points, $\vert \psi_{n\mathbf{k}} \rangle $ is an eigenstate, $\varepsilon_{n\mathbf{k}}$ is an eigenenergy, $\hat{\mathbf{S}} = \frac{\hbar}{2} \hat{\bm{\sigma}}$ is the spin operator, $\hat{\mathbf{v}} = \frac{1}{\hbar}\nabla_\mathbf{k} \hat{H}(\mathbf{k})$ is the velocity operator, $\varepsilon_\mathrm{F}$ is the Fermi level, $f_{n\mathbf{k}}$ is the Fermi--Dirac distribution, and $\Gamma$ is the broadening energy. We use a moderate broadening of $\Gamma = 0.1$~eV.

		We calculate $\delta \mathbf{S}$ per $E_x$ using Eqs.~\eqref{eq:kubo_surf} and \eqref{eq:kubo_sea} for the tight-binding model with $\gamma = -0.05$~eV and $\lambda = 0.02$~eV. By projecting onto sublattices $A$ and $B$, we obtain the sublattice-resolved spin polarizations, denoted by $\delta \mathbf{S}^{A}$ and $\delta \mathbf{S}^{B}$, respectively. Figure~\ref{fig_edelstein}(a) shows $\delta \mathbf{S}^{A,B}/E_x$ as functions of $\varepsilon_\mathrm{F}$ for the conventional antiferromagnetic case with $\Delta=0$~eV. The intraband contribution gives $\delta S_y^{A} = \delta S_y^{B}$, reflecting the isotropic sublattice structure. The interband contribution yields $\delta S_x$ on each sublattice, but $\delta S_x^{A}$ and $\delta S_x^{B}$ are staggered, so the net response vanishes. It is noteworthy that, in noncentrosymmetric antiferromagnets that preserve $\mathcal{PT}$ symmetry~\cite{zelezny2014relativistic, wadley2016electrical, bodnar2018writing, olejnik2018terahertz, salemi2019orbitally}, the sublattice pattern is reversed: the intraband contribution is staggered, whereas the interband contribution is nonstaggered.

		\begin{figure}[t]
			\center\includegraphics[width=1\columnwidth]{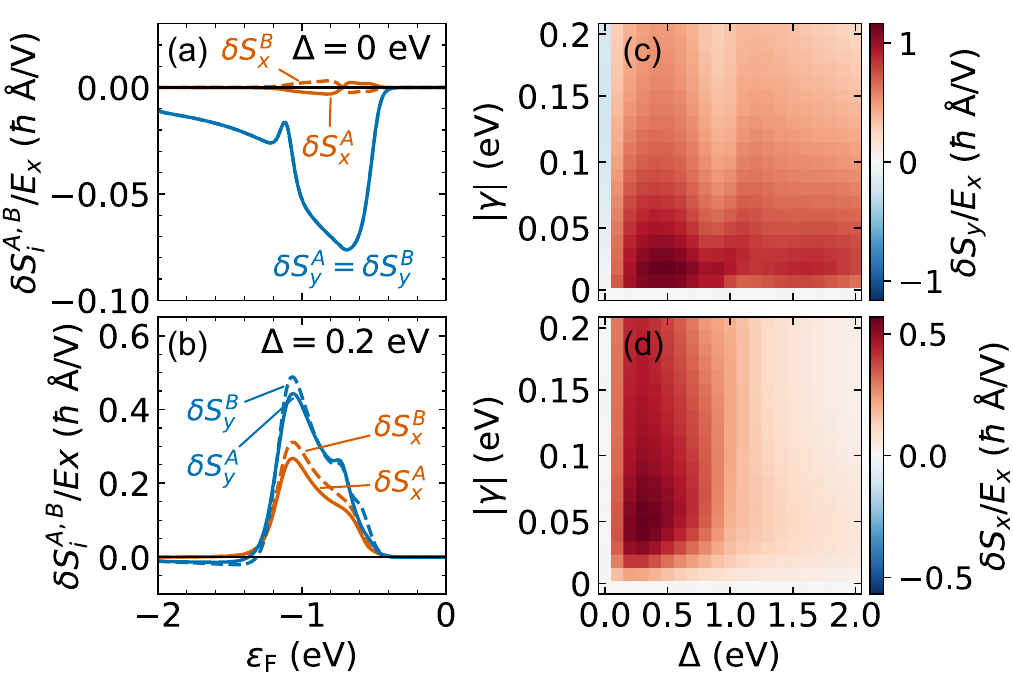}
			\caption{Calculated current-induced spin polarization $\delta S_x$ and $\delta S_y$ per applied electric field $\mathbf{E} = E_x \hat{\mathbf{x}}$. (a),(b) Fermi level ($\varepsilon_\mathrm{F}$) dependences of the sublattice-resolved responses $\delta S_{x}^{A,B}/E_x$ and $\delta S_{y}^{A,B}/E_x$ for systems with (a) $\Delta=0$~eV and (b) $\Delta=0.2$~eV. (c),(d) Net (c) $\delta S_y/E_x$ and (d) $\delta S_x/E_x$ at $\varepsilon_\mathrm{F} = -1$~eV as functions of $\Delta$ and $\gamma$. 
			}
			\label{fig_edelstein} 
		\end{figure}

		Figure~\ref{fig_edelstein}(b) shows the results for the altermagnetic case with $\Delta=0.2$~eV. Here, both $\delta S_y$ and $\delta S_x$ are greatly amplified. Remarkably, neither $\delta S_y$ nor $\delta S_x$ is purely uniform or fully staggered between the two sublattices, reflecting the anisotropic sublattice structure. Such multidirectional spin responses, containing both staggered and nonstaggered components, may enable rich current-induced magnetic dynamics~\cite{manchon2019current}. We note that noncollinear antiferromagnets can also exhibit both intraband and interband spin responses~\cite{gonzalez2024nonrelativistic}, but they originate from noncollinear spin order, so the microscopic origin is different. Our results are also distinct from the Edelstein effects in chiral AMs~\cite{hu2025spin, tenzin2025persistent}, where only the longitudinal component, $\delta {\mathbf{S}} \parallel \mathbf{E}$, arises from the intraband response.

		Figures~\ref{fig_edelstein}(c) and \ref{fig_edelstein}(d) show the net $\delta S_y/E_x$ and $\delta S_x/E_x$ at $\varepsilon_\mathrm{F}=-1.0$~eV, respectively, in the parameter space of $\Delta \ge 0$ and $\gamma \le 0$. For $\Delta = 0$~eV, $\delta S_y$ is finite but not significantly enhanced by increasing |$\gamma$|. Once $\Delta$ becomes finite, however, both components increase markedly even for small values of $\Delta$ and $\gamma$.  From an experimental standpoint, $\gamma$ can be actively tuned via applied gate voltages or strain that breaks inversion symmetry. Given that NRSS often reaches the 0.1--1~eV scale~\cite{bai2024altermagnetism, lee2024broken, fedchenko2024observation, krempasky2024altermagnetic, osumi2024observation}, our results suggest that charge-to-spin conversion can be significantly enhanced in AMs, even under weak SOC. Importantly, such a striking enhancement with $\gamma$ or $\Delta$ is not observed in our ferromagnetic model with momentum-independent exchange splitting~\cite{supp}, which highlights the distinctive role of altermagnetic spin splitting.

		Finally, we discuss the angular dependence of $\delta \mathbf{S}$ on the Néel vector orientation under $\mathbf{E} = E_x \hat{\mathbf{x}}$. We find that both the nonstaggered ($\delta \mathbf{S}^+$) and staggered ($\delta \mathbf{S}^-$) components, defined as $\delta \mathbf{S}^\pm = (\delta \mathbf{S}^A \pm \delta \mathbf{S}^B)/2$, obey the following form~\cite{supp}:
		\begin{equation}\label{eq:angular}
			\delta \mathbf{S}^\pm = E_x \, [  \chi_0^\pm \, \hat{\mathbf{y}} + \chi_1^\pm \, \hat{\mathbf{n}} \times \hat{\mathbf{y}} + \chi_2^\pm \, \hat{\mathbf{n}} \times (\hat{\mathbf{n}} \times \hat{\mathbf{y}}) ].
		\end{equation}
		The $\chi_0^\pm$ and $\chi_2^\pm$ terms together describe the intraband contribution originating from the Rashba spin texture modified by NRSS. The $\hat{\mathbf{n}}$ dependence indicates that the intraband response is sensitive to the relative alignment between the base Rashba spin texture and the NRSS-induced spin polarization along $\hat{\mathbf{n}}$. The $\chi_1^\pm$ term describes the interband or intrinsic spin-torque mechanism, as discussed above. We emphasize that the angular dependence in Eq.~\eqref{eq:angular} is equivalent to that of the spin-orbit torque field in Rashba ferromagnets upon replacing $\hat{\mathbf{n}}$ by the magnetization direction $\hat{\mathbf{m}}$~\cite{li2015intraband}, although the present effect occurs without requiring net magnetization. 
		
		\textit{Material candidates}---Using exploratory first-principles calculations, we find that the proposed mechanism is realized in the noncentrosymmetric $g$-wave AM GdAlSi \cite{nag2024gdalsi, parfenov2025pushing} (see {\hypersetup{linkcolor=black}\hyperlink{mylinkA}{End Matter}}), indicating that it is not restricted to our $d$-wave model but applies more generally to AMs with different spin-splitting symmetries. Other candidates include ferroelectric or polar AMs such as BaCuF$_4$ and Ca$_3$Mn$_2$O$_7$~\cite{gu2025ferroelectric, smejkal2024altermagnetic}, polar altermagnetic phases in strained RuO$_2$ films~\cite{jeong2026altermagnetic}, and altermagnetic thin films or heterostructures in which gating, substrates, or asymmetric interfaces induce ISB. More broadly, our theory suggests that combining NRSS with ISB provides a general strategy for realizing efficient charge-to-spin conversion with weak SOC. 
		
		\textit{Conclusion}---We have shown that Rashba physics in AMs is both qualitatively and quantitatively different from that in conventional Rashba systems. The relativistic SRE emerges from the interplay among the energy scales of NRSS, ISB, and SOC, providing a route to maximizing Rashba spin splitting. Microscopically, the ISB-induced orbital Rashba coupling plays a crucial role by linking NRSS to Rashba spin textures through SOC. The distinctive Rashba Fermi surfaces of AMs enable strongly enhanced charge-to-spin conversion via the SREE. Moreover, Rashba AMs display unconventional Edelstein responses analogous to Rashba spin-orbit torques in ferromagnets, even when net magnetization is absent. 
        
        Because this mechanism relies on the fundamental interplay of these energy scales rather than on a specific orbital geometry, it should be broadly applicable across different altermagnetic symmetries, including both $d$- and $g$-wave systems studied here. These results establish Rashba AMs as platforms for spin-orbitronic applications.

		\textit{Acknowledgments}--- D.J.\ and P.M.O.\ were supported by the Wallenberg Initiative Materials Science for Sustainability (WISE) funded by the Knut and Alice Wallenberg Foundation. P.M.O.\ further acknowledges funding from the Knut and Alice Wallenberg Foundation (Grant No.\ 2022.0079 and No.\ 2023.0336) and the EIC Pathfinder OPEN “OBELIX” (Grant No.\ 101129641). The calculations were supported by resources provided by the National Academic Infrastructure for Supercomputing in Sweden (NAISS) at NSC Linköping, partially funded by Vetenskapsrådet through Grant No.\ 2022-06725 and No.\ 2026-05211. M.\,Y.\ acknowledges the hospitality of Uppsala University during his visit, where part of this work was performed. M.\,Y.\ was supported by the Department of Energy, Office of Basic Energy Sciences, Division of Materials Sciences and Engineering under Contract No.\ DE-FG02-08ER46542 for formal development and manuscript writing.
	}

	\bibliography{ref.bib}
	
	{\allowdisplaybreaks
		\onecolumngrid
		\vspace{0.35cm}
		\subsection{\large End Matter}
		\twocolumngrid

		\hypertarget{mylinkA}{\textit{First-principles calculation for GdAlSi}}---To examine our theory in a real system, we perform first-principles calculations for GdAlSi, which has recently been proposed as a noncentrosymmetric $g$-wave AM~\cite{nag2024gdalsi, parfenov2025pushing}. Figure~\ref{fig_gdalsi}(a) shows the body-centered tetragonal crystal structure of GdAlSi with space group $I4_1md$. The lattice constants are $a=b=4.12$~\r{A} and $c=14.43$~\r{A}, and the atoms occupy the $4a$ Wyckoff positions at $(0, 0, z)$ with $z= 0.374$, 0.958, and 0.793 for Gd, Al, and Si, respectively~\cite{nag2024gdalsi}. The Néel vector is along the $\mathbf{c}$ axis, leading to the magnetic space group $I4'_1m'd$. Notably, the corresponding magnetic point group is $4'm'm$, which is identical to that of our AM model with ISB.

		The electronic structure of GdAlSi is calculated using the density functional theory code \texttt{FLEUR}~\cite{fleurWeb, fleurCode} based on the full-potential linearized augmented plane-wave method~\cite{wimmer1981full}. The Perdew--Burke--Ernzerhof functional~\cite{perdew1996generalized} within the generalized gradient approximation is used for the exchange-correlation functional. A Hubbard $U$~\cite{shick1999implementation} of 7.0~eV~\cite{nag2024gdalsi} is introduced for the Gd 4$f$ electrons. The muffin-tin radii are set to 2.7~$a_0$, 2.15~$a_0$, and 2.0~$a_0$ for Gd, Al, and Si atoms, respectively, where $a_0$ is the Bohr radius, and the plane-wave cutoff is set to 4.5~$a_0^{-1}$. SOC is included within the second variation scheme~\cite{li1990magnetic}. The Brillouin zone for the self-consistent calculation is sampled using a $8 \times 8 \times 8$ Monkhorst--Pack $\mathbf{k}$-mesh~\cite{monkhorst1976special}. 
        Maximally localized Wannier functions~\cite{freimuth2008maximally} are then constructed using the \texttt{WANNIER90} code~\cite{pizzi2020wannier90}, with initial projections of $s$, $p$, $d$, and $f$ orbitals for Gd and $s$ and $p$ orbitals for Al and Si. The frozen- and disentanglement-window energy maxima are set to 5~eV and 15~eV above the Fermi energy, respectively.

		\begin{figure}[t!]
			\center\includegraphics[width=0.95\columnwidth]{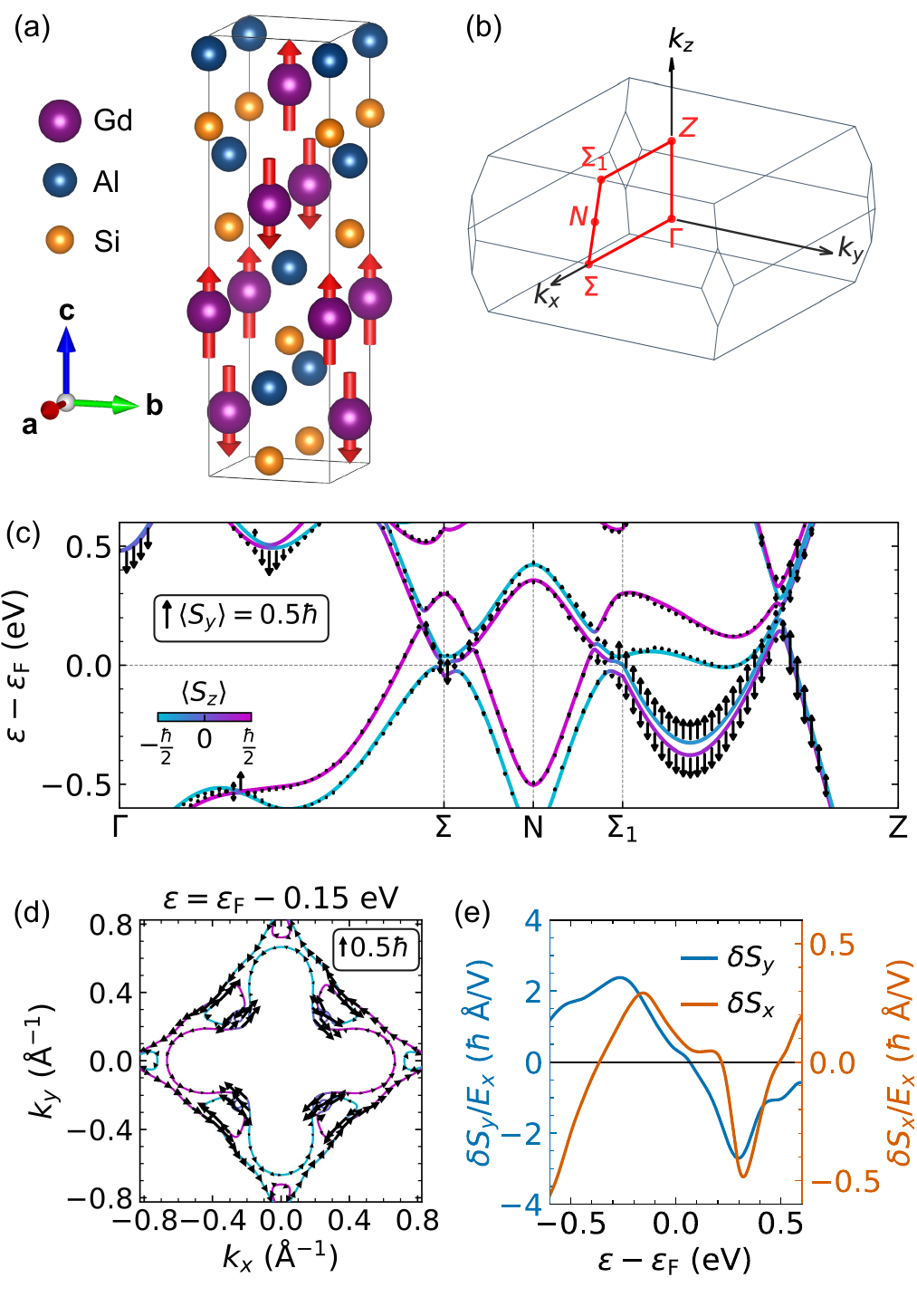}
			\caption{(a) Conventional unit cell of body-centered tetragonal GdAlSi. The lattice vectors $\mathbf{a}$, $\mathbf{b}$, and $\mathbf{c}$ are aligned with the $x$, $y$, and $z$ axes, respectively. Red arrows indicate the spin moments. (b) Brillouin zone of the primitive unit cell of GdAlSi. (c) Band structure obtained from first-principles calculations. Arrows and colors indicate the spin expectation values along the $y$ and $z$ axes, respectively. (d) Spin texture on the Fermi surface in the $k_z=0$ plane at an energy of $\varepsilon_{\mathrm{F}} - 0.15$~eV. (e) Calculated current-induced spin polarization $\delta S_i$ ($i=x,y$) per applied electric field of 1~V/\r{A} along $\hat{\mathbf{x}}$.  
			}
			\label{fig_gdalsi} 
		\end{figure}
		
		The band structure is calculated along the $\mathbf{k}$-path indicated in Fig.~\ref{fig_gdalsi}(b). As shown in Fig.~\ref{fig_gdalsi}(c), the obtained dispersion agrees well with that reported previously~\cite{nag2024gdalsi}. Alongside the out-of-plane spin polarization along $\hat{\mathbf{z}}$ driven by NRSS, the in-plane spin component is firmly locked to $\hat{\mathbf{y}}$. Although GdAlSi is a Weyl semimetal, its broken inversion symmetry can also result in metallic Rashba bands when the Fermi level is slightly shifted, e.g., by doping. Figure~\ref{fig_gdalsi}(d) shows the Fermi surface at an energy of $0.15$~eV below the Fermi level in the $k_z=0$ plane. It clearly demonstrates that the spin-polarized bands host Rashba-type spin textures, with the in-plane spin strongly locked tangentially to the Fermi surface.
		
		Finally, we investigate the current-induced spin polarization, i.e., the spin Edelstein effect, in GdAlSi. Using Wannier interpolation~\cite{pizzi2020wannier90}, we evaluate Eqs.~\eqref{eq:kubo_surf} and \eqref{eq:kubo_sea} on a dense $400 \times 400 \times 400$ $\mathbf{k}$-mesh. The broadening energy is set to $\Gamma=0.01$~eV. This choice is supported by the experimentally measured carrier mobility $\mu \approx 700$ $\mathrm{cm}^{2}\, \mathrm{V}^{-1}\, \mathrm{s}^{-1}$ and effective mass $m^* \approx 0.1 m_e$ at low temperature~\cite{parfenov2025pushing}, where $m_e$ is the free electron mass. These values give a carrier lifetime $\tau = \mu m^*/e \approx 0.04$~ps, determining $\Gamma = \hbar/2\tau$. Similar to our tight-binding model, an applied electric field $\mathbf{E} = E_x \hat{\mathbf{x}}$ generates the conventional SREE component $\delta{S}_y$ and the intrinsic spin-torque component $\delta{S}_x$. Figure~\ref{fig_gdalsi}(e) presents the numerical results. At the Fermi level, we obtain the intraband response $\delta{S}_y/E_x = 0.35 \ \hbar$~\r{A}/V and the interband response $\delta{S}_x/E_x = 0.11 \ \hbar$~\r{A}/V. While these values are already sizable, their energy dependences show that the magnitudes can be further enhanced substantially by tuning the carrier density, reaching magnitudes of approximately $2.5 \ \hbar$~\r{A}/V and $ 0.5 \ \hbar$~\r{A}/V, respectively.

		These values are comparable to or larger than those reported for representative antiferromagnetic spintronic materials. In the noncentrosymmetric antiferromagnets CuMnAs and Mn$_2$Au, the intraband response vanishes after summing over sublattices, while the interband response was estimated to be around 0.05--0.1 $\hbar$~\r{A}/V~\cite{salemi2019orbitally}. In the noncollinear antiferromagnet Mn$_3$Sn, a recent theory~\cite{gonzalez2024nonrelativistic} reported a negligible intraband response at the Fermi level but a comparable interband response on the order of 0.1--0.2 $\hbar$~\r{A}/V, while they can reach the order of 1 $\hbar$~\r{A}/V and 0.5 $\hbar$~\r{A}/V, respectively, upon varying the Fermi level. It is also useful to compare with the SREE in oxide interfaces known for highly efficient charge-to-spin conversion~\cite{trier2022oxide, bihlmayer2022rashba}. For the prototypical two-dimensional electron gas system at the  SrTiO$_3$ interface, a theoretical calculation~\cite{johansson2021spin} predicted a maximum $\delta{S}_y/E_x$ of approximately $5 \ \hbar$~\r{A}/V upon doping. However, it is important to note that the intraband contribution scales with the carrier lifetime $\tau$. In that work, $\tau$ was set to 1~ps---about 25 times larger than the value used in our calculations---hence
        the maximum $\delta{S}_y/E_x$ remains comparable between the two systems. Therefore, GdAlSi provides a promising functionality for altermagnetic spintronics. 
		
	}

	\onecolumngrid
	\clearpage
	
	{\allowdisplaybreaks
		
		\begin{center}
			\textbf{\large \vskip0mm Supplemental Materials for ``Interplay of nonrelativistic and relativistic spin splittings in altermagnets''}
            \vskip3.5mm
			Daegeun Jo,$^{1,2}$ Peter M. Oppeneer$^{1,2}$, and Mohsen Yarmohammadi,$^3$ \vskip1mm
			\small $^1$\textit{Department of Physics and Astronomy, Uppsala University, P.O. Box 516, SE-75120 Uppsala, Sweden}\\
            \small $^2$\textit{Wallenberg Initiative Materials Science for Sustainability, Uppsala University, SE-75120 Uppsala, Sweden}\\
            \small $^3$\textit{Department of Physics, Georgetown University, Washington DC 20057, USA}\\
            (Dated: \today)
		\end{center}

		\setcounter{equation}{0}
		\makeatletter

		\setcounter{equation}{0}
		\renewcommand{\theequation}{S\arabic{equation}}
		\setcounter{figure}{0}
		\renewcommand{\thefigure}{S\arabic{figure}}
		\setcounter{section}{0}
		\renewcommand{\thesection}{S\arabic{section}}
		\setcounter{table}{0}
		\renewcommand{\thetable}{S\arabic{table}}

        {\allowdisplaybreaks
	
	\section{Details of the tight-binding model}\label{sec1}
	
	In the main text, we considered a tight-binding model for a two-dimensional $d$-wave altermagnet (AM). The crystal structure is described by a 45$^\circ$-rotated square lattice with lattice vectors $\mathbf{a}_1 = \frac{a}{\sqrt{2}} \left(1,-1,0 \right)$ and $\mathbf{a}_2 = \frac{a}{\sqrt{2}} \left(1,1,0 \right)$, where we choose $a = 5$~\r{A}. The two sublattices, labeled as $A$ and $B$, are located at $(0, 0, 0)$ and $\frac{1}{2} \left( \mathbf{a}_1 +  \mathbf{a}_2 \right)$, respectively. The orbital basis is chosen as $d_{xy}$, $d_{yz}$, and $d_{zx}$ for each sublattice. We first construct the Hamiltonian without inversion symmetry breaking (ISB). As shown in the main text, the Hamiltonian is expressed in the tensor-product space of orbital, sublattice, and spin degrees of freedom:
	\begin{equation}\label{eq:H_alter}
		\hat{H}_\mathrm{AM}(\mathbf{k}) = \hat{h}(\mathbf{k}) \hat{\tau}_x + \frac{\Delta}{2} (\hat{\ell}_y^2 - \hat{\ell}_x^2) \hat{\tau}_z  - J \hat{\tau}_z  \hat{\mathbf{n}} \cdot \hat{\bm{\sigma}} + \lambda \hat{\bm{\ell}} \cdot \hat{\bm{\sigma}},
	\end{equation}
	where tensor product symbols and identity operators are omitted for brevity. Here, $\tau_i$ and $\sigma_i$ ($i=x,y,z$) are Pauli matrices acting in the sublattice and spin subspaces, respectively. The hopping matrix $\hat{h}(\mathbf{k})$ in the orbital basis, including nearest-neighbor hopping, is given by 
		\begin{equation}
		\hat{h}(\mathbf{k}) = \begin{pmatrix}
			-2t_\pi [\cos(k_xa) + \cos(k_ya)] & 0 & 0 \\
			0 & -2t_\delta \cos(k_xa) - 2t_\pi \cos(k_ya) & 0 \\
			0 & 0 & -2t_\pi \cos(k_xa) - 2t_\delta \cos(k_ya)
		\end{pmatrix}, \label{eq:hopping}
		\end{equation}
	where $t_\pi$ and $t_\delta$ represent the $\pi$- and $\delta$-bonding hopping amplitudes. Without loss of generality, we set $t_\delta = 0$ and $t_\pi  = 0.5$~eV. Since the nearest-neighbor hopping occurs between different sublattices, $\hat{\tau}_x$ appears in the first term on the right-hand side of Eq.~\eqref{eq:H_alter}. The orbital angular momentum (OAM) operators are given by $\hat{L}_i = \hbar \hat{\ell}_i$, where the dimensionless OAM operators are defined as 	
	\begin{equation}
		\hat{\ell}_x = \begin{pmatrix}
			0 & 0 & -i \\
			0 & 0 & 0 \\
			i & 0 & 0
		\end{pmatrix}, \quad
		\hat{\ell}_y = \begin{pmatrix}
			0 & i & 0 \\
			-i & 0 & 0 \\
			0 & 0 & 0
		\end{pmatrix}, \quad
		\hat{\ell}_z = \begin{pmatrix}
			0 & 0 & 0 \\
			0 & 0 & i \\
			0 & -i & 0
		\end{pmatrix}.\label{eq:oam}
	\end{equation}
	The orbital quadrupole term $\hat{\ell}_y^2 - \hat{\ell}_x^2$ in Eq.~\eqref{eq:H_alter} is therefore expressed as
	\begin{equation}
		\hat{\ell}_y^2 - \hat{\ell}_x^2 = \begin{pmatrix}
	0 & 0 & 0 \\
	0 & 1 & 0 \\
	0 &  0 & -1
\end{pmatrix}, \label{eq:quad}
	\end{equation}
	which describes the crystal-field splitting between the $d_{yz}$ and $d_{zx}$ orbitals. Together with $\hat{\tau}_z$, the second term in Eq.~\eqref{eq:H_alter} accounts for the sublattice-dependent crystal-field splitting, reflecting the $d$-wave symmetry. The meanings of the other coefficients and terms are discussed in the main text. 
	
	To incorporate the effect of ISB, we introduce antisymmetric interorbital hopping between the two sublattices, $\langle d_{xy} (\mathbf{0}) \vert \hat{H}_\mathrm{ISB} \vert d_{yz} ( \pm \frac{a}{\sqrt{2}} \hat{\mathbf{x}}  ) \rangle = \langle d_{xy} (\mathbf{0}) \vert \hat{H}_\mathrm{ISB} \vert d_{zx} ( \pm \frac{a}{\sqrt{2}} \hat{\mathbf{y}} ) \rangle \equiv \pm \gamma$, where $\ket{d_{i}(\mathbf{r})}$ denotes the $d_i$ orbital state centered at position $\mathbf{r}$. In the Bloch basis, the Hamiltonian matrix elements are given by
		\begin{equation}
		[\hat{H}_\mathrm{ISB}(\mathbf{k})]_{mn}^{\alpha \beta}  = \sum_\mathbf{R} e^{i \mathbf{k} \cdot (\mathbf{R} + \mathbf{r}_\beta - \mathbf{r}_\alpha)} \bra{d_m^\alpha (\mathbf{r}_\alpha)} \hat{H}_\mathrm{ISB} \ket{d_n^\beta (\mathbf{R}+\mathbf{r}_\beta) }  ,
	\end{equation}
	where $\mathbf{R}$ is a Bravais lattice vector, $m,n$ label orbitals, and $\alpha, \beta = A, B$ label sublattices. Accordingly, we obtain
	\begin{subequations}\begin{align}
		\hat{H}_\mathrm{ISB}(\mathbf{k}) & = - 2  \gamma
				\begin{pmatrix}
			0 & i \sin (\frac{k_x a}{\sqrt{2}}) & i \sin (\frac{k_y a}{\sqrt{2}}) \\
			- i\sin (\frac{k_x a}{\sqrt{2}}) & 0 & 0 \\
			- i\sin (\frac{k_y a}{\sqrt{2}}) & 0 & 0
		\end{pmatrix} 
		\hat{\tau}_x \\
		& = -2 \gamma \left[\sin (\frac{k_x a}{\sqrt{2}}) \hat{\ell}_y - \sin (\frac{k_y a}{\sqrt{2}}) \hat{\ell}_x \right] \hat{\tau}_x. \label{eq:H_ISB}
	\end{align}
    \end{subequations}


	\section{Spin Edelstein effect in the ferromagnetic model}\label{sec2}
	
	To clarify the distinctive role of altermagnetic spin splitting relative to momentum-independent spin splitting in ferromagnets (FMs), we construct a tight-binding FM model that differs from our AM model only in its exchange term. Specifically, the total Hamiltonian is $\hat{H} = \hat{H}_\mathrm{FM} + \hat{H}_\mathrm{ISB}$, where $\hat{H}_\mathrm{ISB}$ is given by Eq.~\eqref{eq:H_ISB}, and
	\begin{equation}\label{eq:H_fm}
		\hat{H}_\mathrm{FM}(\mathbf{k}) = \hat{h}(\mathbf{k}) \hat{\tau}_x + \frac{\Delta}{2} (\hat{\ell}_y^2 - \hat{\ell}_x^2) \hat{\tau}_z  - J \hat{\mathbf{m}} \cdot \hat{\bm{\sigma}} + \lambda \hat{\bm{\ell}} \cdot \hat{\bm{\sigma}}.
	\end{equation}
	This Hamiltonian is equivalent to Eq.~\eqref{eq:H_alter}, with the antiferromagnetic exchange term replaced by a ferromagnetic exchange term. The matrices are defined in Eqs.~\eqref{eq:hopping}--\eqref{eq:quad}. For this model, we evaluate the current-induced spin polarization $\delta \mathbf{S}$ for an applied electric field $\mathbf{E} = E_x \hat{\mathbf{x}}$ using Eqs.~(3a) and (3b) of the main text.
    \begin{figure}[b]
		\center\includegraphics[width=0.5\textwidth]{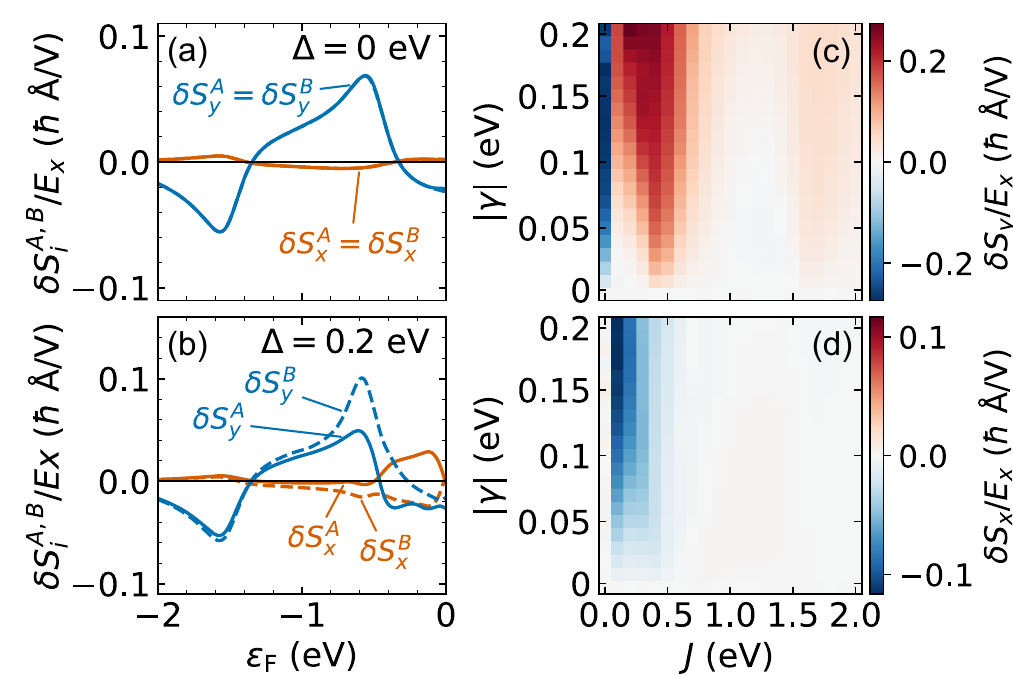}
		\caption{Calculated current-induced spin polarizations $\delta S_x$ and $\delta S_y$ per applied electric field $\mathbf{E} = E_x \hat{\mathbf{x}}$ in the ferromagnetic model. (a),(b) Fermi-level ($\varepsilon_\mathrm{F}$) dependences of the sublattice-resolved responses $\delta S_{x}^{A,B}/E_x$ and $\delta S_{y}^{A,B}/E_x$ for systems with (a) $\Delta=0$~eV and (b) $\Delta=0.2$~eV. (c),(d) Net (c) $\delta S_y/E_x$ and (d) $\delta S_x/E_x$ at $\varepsilon_\mathrm{F} = -0.6$~eV as functions of $J$ and $\gamma$. 
		}
		\label{fig_edelstein_fm} 
	\end{figure}
	
	For comparison with Figs.~4(a) and 4(b) of the main text, we use the same model parameters as for the AM model: $t_\delta=0$, $t_\pi=0.5$~eV, $J=0.5$~eV, $\lambda=0.02$~eV, and $\gamma = -0.05$~eV. The magnetization direction is set to $\hat{\mathbf{m}} = \hat{\mathbf{z}}$. Figures~\ref{fig_edelstein_fm}(a) and \ref{fig_edelstein_fm}(b) show the results for $\Delta = 0$~eV and $\Delta = 0.2$~eV, respectively. We find that the spin responses are not enhanced by $\Delta$, in clear contrast to the pronounced enhancement observed in Figs.~4(a) and (b) of the main text. This behavior is expected because the crystal field is not directly coupled to the ferromagnetic spin splitting, which is governed by $J$. Therefore, the crystal-field enhancement of the spin Edelstein effect mediated by NRSS is a distinctive feature of Rashba AMs. 
	
	In Figs.~4(c) and 4(d) of the main text, we showed that the spin Edelstein effect is significantly enhanced even for relatively weak ISB and NRSS. This enhancement occurs over a broad range of $\Delta$ and $\gamma$. Here, we similarly examine how the spin Edelstein effect depends on the magnitude of the spin splitting in the FM model. Because the ferromagnetic spin splitting is controlled by $J$, we evaluate $\delta S_y/E_x$ and $\delta S_x/E_x$ while varying $J \ge 0$ and $\gamma \le 0$. The Fermi level is set to $\varepsilon_\mathrm{F}=-0.6$~eV, where the responses are large [see, e.g., Fig.~\ref{fig_edelstein_fm}(a)]. As shown in Figs.~\ref{fig_edelstein_fm}(c) and \ref{fig_edelstein_fm}(d), both $\delta S_y/E_x$ and $\delta S_x/E_x$ increase upon introducing a finite $J$, but their overall magnitudes remain smaller than those obtained in the AM model. In particular, their increase with $\gamma$ is relatively gradual, in stark contrast to the strong enhancement observed in the AM model even at $|\gamma| \sim 0.01$--$0.05$~eV over a broad range of $\Delta$.


	\section{Néel vector angular dependence of the spin Edelstein effect}\label{sec3}
	
	We investigate the dependence of the nonequilibrium spin polarization $\delta \mathbf{S}$ on the Néel vector orientation in the Rashba AM model under an applied electric field $\mathbf{E} = E_x \hat{\mathbf{x}}$. For the Hamiltonian $\hat{H}_\mathrm{AM} + \hat{H}_\mathrm{ISB}$, we choose the parameters as $t_\delta=0$, $t_\pi=0.5$~eV, $J=0.5$~eV, $\Delta = 0.5$~eV, $\lambda=0.02$~eV, and $\gamma = -0.05$~eV. We evaluate the nonstaggered ($\delta \mathbf{S}^+$) and staggered ($\delta \mathbf{S}^-$) components, defined by $\delta \mathbf{S}^\pm = (\delta \mathbf{S}^A \pm \delta \mathbf{S}^B)/2$, as functions of the Néel vector orientation $\hat{\mathbf{n}} = (\sin \theta \cos \phi, \sin \theta \sin \phi, \cos \theta) $. As shown in Fig.~\ref{fig_angular}, both $\delta \mathbf{S}^+$ and $\delta \mathbf{S}^-$ follow the angular dependence introduced in the main text:
	\begin{equation}\label{eq:angular}
		\delta \mathbf{S}^\pm = E_x \, [  \chi_0^\pm \, \hat{\mathbf{y}} + \chi_1^\pm \, \hat{\mathbf{n}} \times \hat{\mathbf{y}} + \chi_2^\pm \, \hat{\mathbf{n}} \times (\hat{\mathbf{n}} \times \hat{\mathbf{y}}) ].
	\end{equation}

	\begin{figure}[h]
		\center\includegraphics[width=0.6\textwidth]{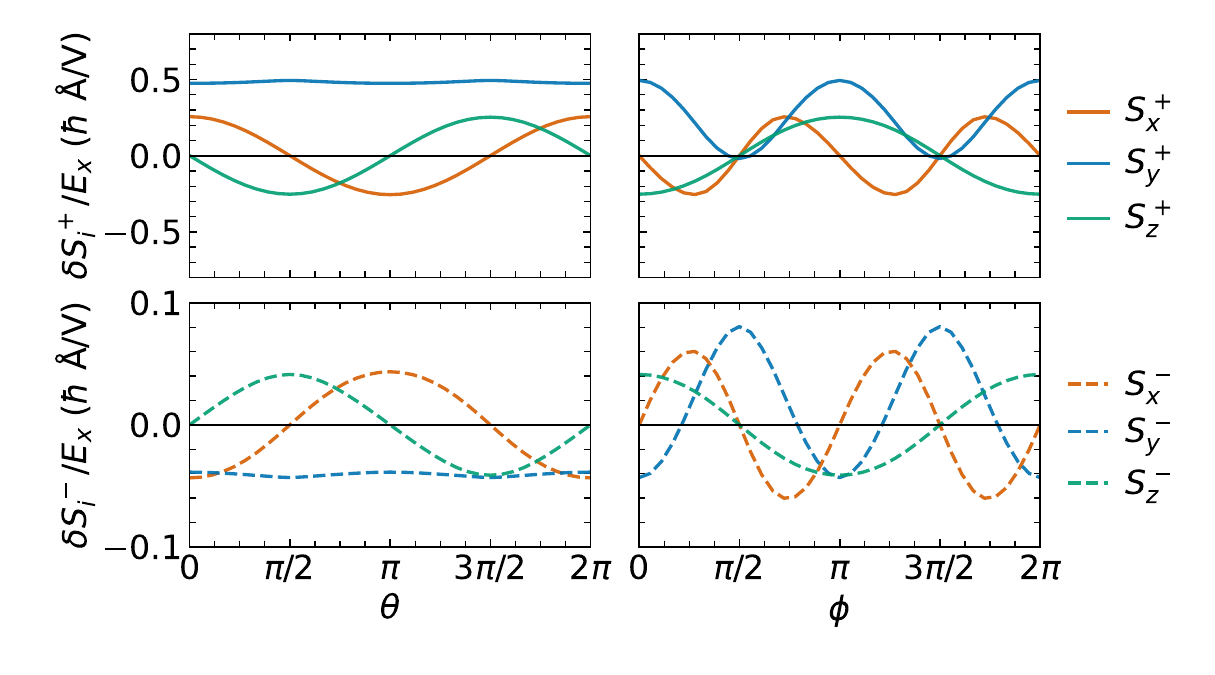}
		\caption{Nonstaggered ($\delta S_i^+$, upper panels) and staggered ($\delta S_i^-$, lower panels) spin polarization components as functions of the Néel vector angle at the Fermi level $\varepsilon_\mathrm{F} = -1$~eV. The polar angle $\theta$ is scanned in the $zx$ plane ($\phi=0$), while the azimuthal angle $\phi$ is scanned in the $xy$ plane ($\theta=\pi/2$).
		}
		\label{fig_angular} 
	\end{figure}

	}
	
\end{document}